# Electrical Detection of Single-Domain Néel Vector Reorientation across the Spin-Flop Transition in $Cr_2O_3$ Crystals

Wei-Cheng Liao[1], Haoyu Liu[1], Weilun Tan[1], Josiah Keagy[1], Jia-mou Chen[1], Jing Shi[1]

[1]*Department of Physics and Astronomy, University of California, Riverside, California 92521, USA*

Abstract

Electrical transport measurements in heterostructures of antiferromagnetic $Cr_2O_3$ bulk crystals and a thin Pt layer exhibit sharp responses as the Néel vector of the $Cr_2O_3$ undergoes the spin-flop transition. This abrupt change can arise from several distinct mechanisms including magnetostriction, proximity-induced anomalous Hall, spin Hall anomalous Hall, and spin Hall planar Hall effects. While large Pt devices sensing multiple up/down domains can produce indistinguishable Hall signal jumps due to different initial Néel vector orientations, smaller Pt devices that sense single domains isolate the proximity-induced Hall signals. This allows direct electrical detection of Néel vector reorientation across the spin-flop transition in single domain regions. Furthermore, the single-domain state can be prepared by magnetic field cooling or magnetoelectric cooling. We demonstrate a method to control and characterize almost the three-dimensional orientation of single-domain Néel vectors by exploiting Hall measurements and cooling techniques, crucial for future antiferromagnetic spintronic applications.

The spin-flop transition (SFT) is an important phenomenon in antiferromagnetic (AFM) materials characterized by a sudden rotation of antiparallel spins, and thus the reorientation of the Néel vector ***N***, accompanied by canting when an applied magnetic field overcomes the magnetic anisotropy. In bulk crystals, the SFT can be detected using various techniques, including neutron scattering[1–4], magnetometry[5–7], magnetic resonance[8–10], and transport methods such as spin Seebeck effect[11–13]. In AFM insulators like $Cr_2O_3$, electrical transport in an adjacent heavy metal layer, such as Pt, can be sensitive to the spin configurations near the interface. This allows for the electrical detection of the Néel vector reorientation at SFT, providing a convenient and versatile interface-sensitive approach particularly suited to AFM thin films and nanodevices. By exploiting this method, we can gain valuable insights into the Néel vector ***N*** at the nanoscale.

In this work, we investigate the Néel vector orientations across the SFT through electrical transport measurements in $Cr_2O_3$/Pt heterostructure devices. We utilize $Cr_2O_3$ single crystals with thicknesses of 0.5 mm and 1.0 mm. Two crystal orientations are chosen: $(10\bar{1}0)$ and $(0001)$. In the $(10\bar{1}0)$-oriented crystal, the $c$-axis lies in the surface plane, resulting in a compensated surface. Conversely, in the $(0001)$-oriented crystal, the $c$-axis is perpendicular to the surface, and the surface spins of a single domain originate from only one AFM sublattice, creating an uncompensated surface. The polished $Cr_2O_3$ crystal surface is smooth with a root-mean-square (RMS) roughness of approximately 0.2 nm, which is subsequently coated with a 3 or 5 nm thick Pt layer using magnetron sputtering. Photolithography or electron beam lithography is used to define Pt Hall bar or Hall cross patterns with lateral dimensions ranging from 100 μm to 0.1 μm (see Supporting Information). Transport measurements on these devices are performed using Quantum Design's Physical Property Measurement System.

We start with the $(10\bar{1}0)$-orientation with a compensated surface of $Cr_2O_3$ on which no intentional net magnetic moments participate in the electrical transport of Pt placed on top. As the in-plane magnetic field is swept along the $c$-axis, the uniaxial direction of $Cr_2O_3$, there is an abrupt jump at 6 T, the spin-flop field, in the spin Seebeck effect (SSE) response as shown in Figure 1(a), signaling the switching of magnon spin polarization. More details about the SSE features were reported previously[12,13]. This SSE jump marks the discontinuity in spin-wave excitation spectrum in $Cr_2O_3$. Since the near-90° sudden rotation of static spin configuration occurs at this transition, it is expected to be sensed resistively using Pt[14]. A clear upward jump appears in the longitudinal magnetoresistance (MR) of Pt, $\frac{\Delta R}{R}$, at approximately the same field (Figure 1(b)). It is tempting to assign this MR jump to the spin-Hall magnetoresistance (SMR) effect just as in bilayers consisting of a ferrimagnetic insulator and a heavy metals[15–18]. However, a closer examination reveals crucial discrepancies. First, as Néel vector ***N*** rotates away from the *c*-axis, i.e., the current flow direction in the device, it should give rise to a downward SMR jump, opposite to the trend observed here. Second, to eliminate the SMR effect, we insert a 10 nm-thick, pinhole-free $Al_2O_3$ layer between $Cr_2O_3$ and Pt. Despite this insulating layer, we still observe a similar magnetoresistance upward jump, but with a slightly reduced magnitude (Figure 1(c)). These facts suggest that a spin-independent magnetoresistance mechanism dominates the SMR effect in $Cr_2O_3$/Pt. It was known that the magnetostriction effect causes the material to elongate along the *c*-axis while contracting laterally upon the spin-flop transition[19,20]. We find that this effect can quantitatively account for a resistance jump of $\frac{\Delta R}{R} = \frac{2\Delta L}{L} \sim 5.6 \times 10^{-5}$ at 20 K, which is close to our observed value of $4.5 \times 10^{-5}$ in $Cr_2O_3$/Pt at 20 K[19,20]. The subtle difference between these two values might be attributed to the SMR effect. Evidently, the magnetostriction-dominated longitudinal magnetoresistance lacks sufficient sensitivity required to probe the spin configurations.

To electrically detect the orientation of ***N*** more decisively, we exploit the Hall effect which is inherently insensitive to the resistance change caused by a magnetostriction effect. By utilizing the uncompensated (0001) surface of $Cr_2O_3$, we induce a net magnetic moment ***m*** in Pt via the proximity effect. For $Cr_2O_3$, even above the SFT, the field-induced canting is very small due to the strong exchange interaction relative to the anisotropy. Therefore, the orientation of one sublattice spins can approximately represent that of ***N***, which means that the direction of the induced ***m*** in Pt can serve as a reliable indicator of the direction of ***N***. As ***N*** reorients, the change in the Hall signal can be generated through two mechanisms: the anomalous Hall effect (AHE) arising from $m_z$, the perpendicular component of ***m***; and the planar Hall effect (PHE)[6,21,22]. We acknowledge that the AHE signals in $Cr_2O_3$ thin films have been attributed to the proximity-induced magnetic layer at the interface[23,24] or the spin current reflection by the interfacial moments[25,26], both of which is proportional to $m_z$. The PHE is proportional to either $m_x m_y$ (due to the in-plane components of ***m*** in Pt), or $N_x N_y$ (due to the in-plane components of ***N*** via the spin current generated by the bulk $Cr_2O_3$), which provides information about the in-plane orientation of ***m*** or ***N***. Unlike PHE, AHE can detect whether the 180° reversal occurs for ***m*** or ***N***. We perform Hall measurements on a 50 μm wide $Cr_2O_3$/Pt Hall cross device by passing a current through each of the two orthogonal channels in geometries G and G', as illustrated in Figure 2(a). The magnetic field is intentionally tilted slightly away from the *c*-axis, with an in-plane component approximately aligned with the $\varphi = 45°$ direction to maximize the PHE responses above the SFT. As the magnetic field is swept, the transverse resistances $R_{yx}$ in both G and G' are measured, as shown in the inset of Figure 2(b). After subtracting the linear ordinary Hall effect (OHE) background, the corrected Hall signals are presented in the main panel of Figure 2(b). In each Hall channel, sharp jumps occur at ±6 T, the SFT fields where ***m*** and ***N*** abruptly rotate by approximately 90° from out-of-plane to in-plane. Above the SFT, the AHE signal is expected to vanish due to the zero $m_z$ component, but the PHE signal can suddenly jump from zero to a finite value which is proportional to $\sin 2\varphi$. In theory, both the z-component and the in-plane component of ***m*** and ***N*** are obtained from the AHE and PHE responses, subject to an ambiguity in the in-plane direction between $\varphi$ and $\varphi + \pi$.[6] Notably, the two orthogonal channels in G and G' report opposite PHE signals if there is an in-plane ***m*** or ***N*** component.

By averaging the Hall signals from G and G', i.e., $(R_H(G) + R_H(G'))/2$, we effectively eliminate both the mixed signal from the longitudinal channel due to unintended device geometry asymmetry[27,28] and the PHE signals as discussed above, leaving only the AHE signal, $R_{AHE}$. Conversely, the difference signal contains only the PHE and geometry-induced mixed signals, which produce opposite signs in G and G' channels. The well-defined Hall cross devices result in negligible mixing of the longitudinal signal (see Supporting Information); therefore, we attribute the half of the difference, i.e., $(R_H(G) - R_H(G'))/2$, to the PHE signal $R_{PHE}$. Figure 2(c) displays both $R_{AHE}$ and $R_{PHE}$ curves. Clearly, the $R_{AHE}$ signal remains nearly flat across the entire field range, showing no observable features at the SFT. The absence of $R_{AHE}$ below the SFT suggests the presence of multiple domains within the 50 μm ×50 μm region, leading to a net <$m_z$> of approximately zero. On the other hand, because the $R_{PHE}$ signal cannot distinguish between the two oppositely aligned in-plane directions of ***N***, the rotation of ***N*** from the initial multi-domain state to the in-plane direction does not result in signal cancellation even if there are spin-flop domains, and therefore produces finite $R_{PHE}$ jumps.

While we have demonstrated the sensitivity of the PHE responses to Néel vector rotation at the SFT, the absence of the AHE signal, resulting from the cancellation of $m_z$ in multiple domains, hinders the determination of the state of the Néel vector below the SFT. To unequivocally detect the proximity-induced AHE signal in Pt below SFT, smaller Hall cross devices capable of probing single domains are necessary. We fabricate smaller Pt Hall cross devices with widths ranging from 10 μm to 0.1 µm using electron beam lithography. Figure 3(a) illustrates the Hall signals, $R_H(G)$, and $R_H(G')$, obtained at 2 K from the G and G' channels of a 10 µm device subjected to zero magnetic field (ZFC) from 320 K, exceeding the Néel temperature ($T_N \sim 307$ K). Both $R_H(G)$ and $R_H(G')$ signals exhibit a sharp SFT, similar to what was observed in the 50 µm device. However, in contrast to the 50 µm device, wherein $R_H(G)$ and $R_H(G')$ display opposing jump directions, the 10 µm device demonstrates concurrent downward jumps in both signals at the SFT fields. This results in a non-zero $R_{AHE}$ signal below the SFT. This observed finite $R_{AHE}$ is attributed to a net $m_z$ within the 10 µm × 10 µm area, arising from either incomplete $m_z$ cancellation due to multi-domain configurations or a single domain state. This behavior is in marked contrast to that of the 50 µm device.

To influence the ZFC state, field cooling (FC) is performed from 320 K to 2 K under positive (PFC) and negative (NFC) 1 T fields. The corresponding Hall responses, presented in Figures 3(b) and 3(c), demonstrate a near inversion of the $R_H(G)$ and $R_H(G')$ features between PFC and NFC, which suggests opposite signs of $m_z$ resulting from the application of opposing FC fields. Figures 3(d), 3(e), and 3(f) display the $R_{AHE}$ and $R_{PHE}$ curves for ZFC, PFC, and NFC, respectively. While $R_{PHE}$ curves remain largely unchanged and are approximately field symmetric, $R_{AHE}$ curves display distinct differences. First, the jumps at the SFT fields can be upwards or downwards, depending on FC conditions. Second, disregarding the difference in SFT jump sizes, the overall PFC and NFC responses are inverted. Third, the ZFC curve closely resembles the NFC curve. The three $R_{AHE}$ curves are plotted together in Figure 3(g). Notably, while the NFC and PFC conditions produce two distinct AHE values below the SFT, the two curves converge above the SFT. Moreover, the fact that the ZFC curve is nearly identical to the NFC curve suggests that even the ZFC state is a single-domain state. These results collectively demonstrate that the Néel vector of the single domain can be reversed solely by a cooling magnetic field, which can be electrically detected. While the detailed mechanism by which FC influences AFM domains remains unclear, similar FC effects have been previously reported by other groups[23,29–33]. Such FC effects were tentatively attributed[29] to an above-Néel-temperature memory effect within residual AFM domain walls, a phenomenon that warrants further investigation. It is crucial to observe that, despite the main curve's inversion indicating the reversal of **N**, the features at the SFT fields are not symmetric about the field. In fact, the two FC $R_{AHE}$ curves are not perfectly inverted. By averaging the two FC $R_{AHE}$ curves shown in Figures 3(e) and 3(f), we can extract the non-inverted component as depicted in Figure 3(h). The field-antisymmetric component, $R_{AHE\_A}$, invariant with cooling field direction, is removed from Figure 3(g). This isolates the cooling field-reversible, field-symmetric components, $R_{AHE\_S}$, as shown in Figure 3(i). FC process establishes **m** and **N** in one of two single-domain states, indicated by the two plateaus on the $R_{AHE\_S}$ curve. During successive low-temperature field sweeps between ±14 T, the $R_{AHE\_S}$ curve repeats, demonstrating that **m** and **N**, once in a single-domain state, maintain their orientation below the SFT and undergo reversible rotation across the SFT, regardless of sweeping field direction. These behaviors are qualitatively different from those reported in thin films where the ferromagnetic-like AHE loops are observed[23,24]. In contrast, the $R_{AHE\_A}$ component exhibits a distinct behavior, changing sign with field reversal. Intriguingly, the $R_{AHE\_A}$ component peaks near the SFT fields and diminishes with increasing field, where **m** and **N** are expected to lie within the film plane. However, in-plane rotation of **m** and **N** would not produce an averaged Hall signal. Therefore, we suggest alternative mechanisms, such as the topological Hall effect[34–36], induced by spin textures that form when the effective anisotropy weakens near the SFT.

We have demonstrated that the cooling field direction can reliably toggle the Néel vector state with the single-domain region under the 10 μm Pt Hall cross device. To validate this FC effect, we have performed similar FC experiments on four additional Hall cross devices, fabricated on a separate $Cr_2O_3$ crystal using electron beam lithography and Pt etching (see Supporting Information). These devices, A and B (10 μm), C (1 μm), and D (0.1 μm) in width, are shown in Figure 4. The upper two rows present the $R_{AHE}$ results for all four devices under 1 T positive (PFC) and negative (NPC) fields, with arrows indicating the corresponding Néel vector orientations. Notably, all devices, regardless of size, exhibit nearly identical $R_{AHE}$ plateau magnitudes (~ 3 $m\Omega$), confirming that they sense single domains. Furthermore, the Néel vector is consistently reversed between PFC and NPC, reinforcing our previous findings. Interestingly, despite being located up to approximately 0.8 mm from each other on the same substrate and subjected to the identical cooling field, the Néel vector orientation is not uniform across the substrate. These findings indicate that while FC is effective in switching Néel vector orientations, it is not sufficient for the deterministic selection of a specific orientation to achieve a single AFM domain across the entire material.

$Cr_2O_3$ is a magnetoelectric material where the Néel vector can be controlled by both electric and magnetic fields[29–31,37–49]. Utilizing Pt Hall crosses as the top electrode and silver paint on the bottom surface as the bottom electrode, we demonstrate deterministic Néel vector orientation setting via magnetoelectric cooling (MEC). The lower two rows of Figure 4 show the $R_{AHE}$ results for three devices subjected to MEC with a 1 T magnetic field and ±50 V across the 0.5 mm thick $Cr_2O_3$ crystal, corresponding to an electric field of 1 kV/cm. With +50 V (PMEC), the Néel vector is consistently set downward in three tested devices. Conversely, a -50 V (NMEC) results in an upward Néel vector direction in the same devices, independent of their size. This confirms the deterministic control of Néel vector orientation using MEC.

Both the PHE and proximity-induced AHE signals in a single domain should be proportional to the magnitude of the Néel vector and other relevant parameters such as the spin-mixing conductance and anomalous Hall angle. The temperature dependence of these two signals is measured and shown in Figure 5 for Device A in Figure 4. As the Néel temperature is approached, both signals decrease as expected. The overall temperature dependence exhibits a broad maximum around 100 K. A similar non-monotonic response was observed in the magneto-optical Kerr effect signal from the top $Cr_2O_3$ surface[44]. The broad peaks in both AHE and PHE likely originate from the $Cr_2O_3$(0001) sublattice moment and therefore the magnitude of the Néel vector, both of which are imprinted in the Hall responses of Pt.

In conclusion, we present the unequivocal electrical detection of single-domain Néel vector reorientation across the SFT of $Cr_2O_3$, using micro- and nano-fabricated Pt Hall cross devices. Magnetoelectric cooling provides deterministic control of the single-domain Néel vector, unlike the domain configuration changes induced by magnetic field cooling. This precise manipulation and electrical readout of the antiferromagnetic (AFM) order parameter at the single-domain level represents a key milestone in the development of AFM spintronics.

**Supporting Information**

- Additional experimental details about device fabrication and characterization, measurement techniques used, analysis of planar Hall effect distinct from geometry-induced mixing, decomposition of the Hall signal at spin-flop transition, and temperature dependence of the antisymmetric part of the anomalous Hall effect.

We thank Víctor Ortiz, Xinping Shi and Richard Wilson for their help with sputtering and Quanshui Xu and Igor Barsukov for their useful discussions. We acknowledge the support provided

References:

(1) Lauter-Pasyuk, V.; Lauter, H. J.; Toperverg, B. P.; Romashev, L.; Ustinov, V. Transverse and Lateral Structure of the Spin-Flop Phase in Fe/Cr Antiferromagnetic Superlattices. *Phys. Rev. Lett.* **2002**, *89* (16), 167203. https://doi.org/10.1103/PhysRevLett.89.167203.

(2) Zheludev, A.; Ressouche, E.; Tsukada, I.; Masuda, T.; Uchinokura, K. Structure of Multiple Spin-Flop States in $BaCu_2Si_2O_7$. *Phys. Rev. B* **2002**, *65* (17), 174416. https://doi.org/10.1103/PhysRevB.65.174416.

(3) Qureshi, N.; Valldor, M.; Weber, L.; Senyshyn, A.; Sidis, Y.; Braden, M. Magnetic Spin-Flop Transition and Interlayer Spin-Wave Dispersion in $PrCaFeO_4$ Revealed by Neutron Diffraction and Inelastic Neutron Scattering. *Phys. Rev. B* **2015**, *91* (22), 224402. https://doi.org/10.1103/PhysRevB.91.224402.

(4) Gitgeatpong, G.; Zhao, Y.; Piyawongwatthana, P.; Qiu, Y.; Harriger, L. W.; Butch, N. P.; Sato, T. J.; Matan, K. Nonreciprocal Magnons and Symmetry-Breaking in the Noncentrosymmetric Antiferromagnet. *Phys. Rev. Lett.* **2017**, *119* (4), 047201. https://doi.org/10.1103/PhysRevLett.119.047201.

(5) Wang, M.; Andrews, C.; Reimers, S.; Amin, O. J.; Wadley, P.; Campion, R. P.; Poole, S. F.; Felton, J.; Edmonds, K. W.; Gallagher, B. L.; Rushforth, A. W.; Makarovsky, O.; Gas, K.; Sawicki, M.; Kriegner, D.; Zubáč, J.; Olejník, K.; Novák, V.; Jungwirth, T.; Shahrokhvand, M.; Zeitler, U.; Dhesi, S. S.; Maccherozzi, F. Spin Flop and Crystalline Anisotropic Magnetoresistance in CuMnAs. *Phys. Rev. B* **2020**, *101* (9), 094429. https://doi.org/10.1103/PhysRevB.101.094429.

(6) Zhang, Y.-H.; Chuang, T.-C.; Qu, D.; Huang, S.-Y. Detection and Manipulation of the Antiferromagnetic Néel Vector in $Cr_2O_3$. *Phys. Rev. B* **2022**, *105* (9), 094442. https://doi.org/10.1103/PhysRevB.105.094442.

(7) Nogués, J.; Morellon, L.; Leighton, C.; Ibarra, M. R.; Schuller, I. K. Antiferromagnetic Spin Flop and Exchange Bias. *Phys. Rev. B* **2000**, *61* (10), R6455–R6458. https://doi.org/10.1103/PhysRevB.61.R6455.

(8) Foner, S. High-Field Antiferromagnetic Resonance in $Cr_2O_3$. *Phys. Rev.* **1963**, *130* (1), 183–197. https://doi.org/10.1103/PhysRev.130.183.

(9) Li, J.; Wilson, C. B.; Cheng, R.; Lohmann, M.; Kavand, M.; Yuan, W.; Aldosary, M.; Agladze, N.; Wei, P.; Sherwin, M. S.; Shi, J. Spin Current from Sub-Terahertz-Generated Antiferromagnetic Magnons. *Nature* **2020**, *578* (7793), 70–74. https://doi.org/10.1038/s41586-020-1950-4.

(10) Yuan, W.; Li, J.; Shi, J. Spin Current Generation and Detection in Uniaxial Antiferromagnetic Insulators. *Appl. Phys. Lett.* **2020**, *117* (10). https://doi.org/10.1063/5.0022391.

(11) Wu, S. M.; Zhang, W.; KC, A.; Borisov, P.; Pearson, J. E.; Jiang, J. S.; Lederman, D.; Hoffmann, A.; Bhattacharya, A. Antiferromagnetic Spin Seebeck Effect. *Phys. Rev. Lett.* **2016**, *116* (9), 097204. https://doi.org/10.1103/PhysRevLett.116.097204.

(12) Reitz, D.; Li, J.; Yuan, W.; Shi, J.; Tserkovnyak, Y. Spin Seebeck Effect Near the Antiferromagnetic Spin-Flop Transition. *Phys. Rev. B* **2020**, *102* (2), 020408. https://doi.org/10.1103/PhysRevB.102.020408.

(13) Li, J.; Simensen, H. T.; Reitz, D.; Sun, Q.; Yuan, W.; Li, C.; Tserkovnyak, Y.; Brataas, A.; Shi, J. Observation of Magnon Polarons in a Uniaxial Antiferromagnetic Insulator. *Phys. Rev. Lett.* **2020**, *125* (21), 217201. https://doi.org/10.1103/PhysRevLett.125.217201.

(14) Su, T.; Lohmann, M.; Li, J.; Xu, Y.; Niu, B.; Alghamdi, M.; Zhou, H.; Cui, Y.; Cheng, R.; Taniguchi, T.; Watanabe, K.; Shi, J. Current-Induced $CrI_3$ Surface Spin-Flop Transition Probed by Proximity Magnetoresistance in Pt. *2d Mater.* **2020**, *7* (4), 045006. https://doi.org/10.1088/2053-1583/ab9dd5.

(15) Nakayama, H.; Althammer, M.; Chen, Y.-T.; Uchida, K.; Kajiwara, Y.; Kikuchi, D.; Ohtani, T.; Geprägs, S.; Opel, M.; Takahashi, S.; Gross, R.; Bauer, G. E. W.; Goennenwein, S. T. B.; Saitoh, E. Spin Hall Magnetoresistance Induced by a Nonequilibrium Proximity Effect. *Phys. Rev. Lett.* **2013**, *110* (20), 206601. https://doi.org/10.1103/PhysRevLett.110.206601.

(16) Chen, Y.-T.; Takahashi, S.; Nakayama, H.; Althammer, M.; Goennenwein, S. T. B.; Saitoh, E.; Bauer, G. E. W. Theory of Spin Hall Magnetoresistance. *Phys. Rev. B* **2013**, *87* (14), 144411. https://doi.org/10.1103/PhysRevB.87.144411.

(17) Althammer, M.; Meyer, S.; Nakayama, H.; Schreier, M.; Altmannshofer, S.; Weiler, M.; Huebl, H.; Geprägs, S.; Opel, M.; Gross, R.; Meier, D.; Klewe, C.; Kuschel, T.; Schmalhorst, J.-M.; Reiss, G.; Shen, L.; Gupta, A.; Chen, Y.-T.; Bauer, G. E. W.; Saitoh, E.; Goennenwein, S. T. B. Quantitative Study of the Spin Hall Magnetoresistance in Ferromagnetic Insulator/Normal Metal Hybrids. *Phys. Rev. B* **2013**, *87* (22), 224401. https://doi.org/10.1103/PhysRevB.87.224401.

(18) Vlietstra, N.; Shan, J.; Castel, V.; van Wees, B. J.; Ben Youssef, J. Spin-Hall Magnetoresistance in Platinum on Yttrium Iron Garnet: Dependence on Platinum Thickness and in-Plane/out-of-Plane Magnetization. *Phys. Rev. B* **2013**, *87* (18), 184421. https://doi.org/10.1103/PhysRevB.87.184421.

(19) Dudko, K. L.; Eremenko, V. V.; Semenenko, L. M. Magnetostriction of Antiferromagnetic $Cr_2O_3$ in Strong Magnetic Fields. *Phys. Status Solidi B Basic Res.* **1971**, *43* (2), 471–477. https://doi.org/10.1002/pssb.2220430203.

(20) Eremenko, V. V.; Sirenko, V. A. Magnetostriction and Spin-Flopping of Uniaxially Compressed Antiferromagnets. In *Modern Trends in Magnetostriction Study and Application*; Springer Netherlands: Dordrecht, 2001; pp 223–247. https://doi.org/10.1007/978-94-010-0959-1_11.

(21) Fischer, J.; Gomonay, O.; Schlitz, R.; Ganzhorn, K.; Vlietstra, N.; Althammer, M.; Huebl, H.; Opel, M.; Gross, R.; Goennenwein, S. T. B.; Geprägs, S. Spin Hall Magnetoresistance in Antiferromagnet/Heavy-Metal Heterostructures. *Phys. Rev. B* **2018**, *97* (1), 014417. https://doi.org/10.1103/PhysRevB.97.014417.

(22) Mahmood, A.; Echtenkamp, W.; Street, M.; Wang, J.-L.; Cao, S.; Komesu, T.; Dowben, P. A.; Buragohain, P.; Lu, H.; Gruverman, A.; Parthasarathy, A.; Rakheja, S.; Binek, C. Voltage Controlled Néel Vector Rotation in Zero Magnetic Field. *Nat. Commun.* **2021**, *12* (1), 1674. https://doi.org/10.1038/s41467-021-21872-3.

(23) Ujimoto, K.; Sameshima, H.; Toyoki, K.; Kotani, Y.; Moriyama, T.; Nakamura, K.; Nakatani, R.; Shiratsuchi, Y. Direct Observation of Antiferromagnetic Domains and Field-Induced Reversal in $Pt/Cr_2O_3/Pt$ Epitaxial Trilayers. *Appl. Phys. Lett.* **2023**, *123* (2). https://doi.org/10.1063/5.0156254.

(24) Moriyama, T.; Shiratsuchi, Y.; Iino, T.; Aono, H.; Suzuki, M.; Nakamura, T.; Kotani, Y.; Nakatani, R.; Nakamura, K.; Ono, T. Giant Anomalous Hall Conductivity at the $Pt/Cr_2O_3$ Interface. *Phys. Rev. Appl.* **2020**, *13* (3), 034052. https://doi.org/10.1103/PhysRevApplied.13.034052.

(25) Ji, Y.; Miao, J.; Meng, K. K.; Ren, Z. Y.; Dong, B. W.; Xu, X. G.; Wu, Y.; Jiang, Y. Spin Hall Magnetoresistance in an Antiferromagnetic Magnetoelectric $Cr_2O_3$/Heavy-Metal W Heterostructure. *Appl. Phys. Lett.* **2017**, *110* (26), 262401. https://doi.org/10.1063/1.4989680.

(26) Schlitz, R.; Kosub, T.; Thomas, A.; Fabretti, S.; Nielsch, K.; Makarov, D.; Goennenwein, S. T. B. Evolution of the Spin Hall Magnetoresistance in $Cr_2O_3$/Pt Bilayers Close to the Néel Temperature. *Appl. Phys. Lett.* **2018**, *112* (13), 132401. https://doi.org/10.1063/1.5019934.

(27) Kosub, T.; Kopte, M.; Radu, F.; Schmidt, O. G.; Makarov, D. All-Electric Access to the Magnetic-Field-Invariant Magnetization of Antiferromagnets. *Phys. Rev. Lett.* **2015**, *115* (9), 097201. https://doi.org/10.1103/PhysRevLett.115.097201.

(28) Daniil, P.; Cohen, E. Low Field Hall Effect Magnetometry. *J. Appl. Phys.* **1982**, *53* (11), 8257–8259. https://doi.org/10.1063/1.330300.

(29) Du, K.; Xu, X.; Won, C.; Wang, K.; Crooker, S. A.; Rangan, S.; Bartynski, R.; Cheong, S.-W. Topological Surface Magnetism and Néel Vector Control in a Magnetoelectric Antiferromagnet. *npj Quantum Mater.* **2023**, *8* (1), 17. https://doi.org/10.1038/s41535-023-00551-0.

(30) Kosub, T.; Kopte, M.; Hühne, R.; Appel, P.; Shields, B.; Maletinsky, P.; Hübner, R.; Liedke, M. O.; Fassbender, J.; Schmidt, O. G.; Makarov, D. Purely Antiferromagnetic Magnetoelectric Random Access Memory. *Nat. Commun.* **2017**, *8* (1), 13985. https://doi.org/10.1038/ncomms13985.

(31) Fallarino, L.; Berger, A.; Binek, C. Magnetic Field Induced Switching of the Antiferromagnetic Order Parameter in Thin Films of Magnetoelectric Chromia. *Phys. Rev. B* **2015**, *91* (5), 054414. https://doi.org/10.1103/PhysRevB.91.054414.

(32) Pylypovskyi, O. V.; Weber, S. F.; Makushko, P.; Veremchuk, I.; Spaldin, N. A.; Makarov, D. Surface-Symmetry-Driven Dzyaloshinskii-Moriya Interaction and Canted Ferrimagnetism in Collinear Magnetoelectric Antiferromagnet $Cr_2O_3$. *Phys. Rev. Lett.* **2024**, *132* (22), 226702. https://doi.org/10.1103/PhysRevLett.132.226702.

(33) Brown, P. J.; Forsyth, J. B.; Tasset, F. A Study of Magnetoelectric Domain Formation in $Cr_2O_3$. *J. Phys. Condens. Matter* **1998**, *10* (3), 663–672. https://doi.org/10.1088/0953-8984/10/3/017.

(34) Neubauer, A.; Pfleiderer, C.; Binz, B.; Rosch, A.; Ritz, R.; Niklowitz, P. G.; Böni, P. Topological Hall Effect in the α Phase of MnSi. *Phys. Rev. Lett.* **2009**, *102* (18), 186602. https://doi.org/10.1103/PhysRevLett.102.186602.

(35) Sürgers, C.; Fischer, G.; Winkel, P.; Löhneysen, H. v. Large Topological Hall Effect in the Non-Collinear Phase of an Antiferromagnet. *Nat. Commun.* **2014**, *5* (1), 3400. https://doi.org/10.1038/ncomms4400.

(36) Shao, Q.; Liu, Y.; Yu, G.; Kim, S. K.; Che, X.; Tang, C.; He, Q. L.; Tserkovnyak, Y.; Shi, J.; Wang, K. L. Topological Hall Effect at above Room Temperature in Heterostructures Composed of a Magnetic Insulator and a Heavy Metal. *Nat. Electron.* **2019**, *2* (5), 182–186. https://doi.org/10.1038/s41928-019-0246-x.

(37) Dzyaloshinskii, I. E. On the Magneto-Electrical Effects in Antiferromagnets. *Sov. Phys. JETP* **1960**, *10*, 628–629.

(38) Astrov, D. N. The Magnetoelectric Effect in Antiferromagnetics. *Sov. Phys. JETP* **1960**, *11* (3), 708–709.

(39) Astrov, D. N. Magnetoelectric Effect in Chromium Oxide. *Sov. Phys. JETP* **1961**, *13* (4), 729–733.

(40) Folen, V. J.; Rado, G. T.; Stalder, E. W. Anisotropy of the Magnetoelectric Effect in $Cr_2O_3$. *Phys. Rev. Lett.* **1961**, *6* (11), 607–608. https://doi.org/10.1103/PhysRevLett.6.607.

(41) Rado, G. T.; Folen, V. J. Observation of the Magnetically Induced Magnetoelectric Effect and Evidence for Antiferromagnetic Domains. *Phys. Rev. Lett.* **1961**, *7* (8), 310–311. https://doi.org/10.1103/PhysRevLett.7.310.

(42) Martin, T.; Anderson, J. Antiferromagnetic Domain Switching in $Cr_2O_3$. *IEEE Trans. Magn.* **1966**, *2* (3), 446–449. https://doi.org/10.1109/TMAG.1966.1065857.

(43) Borisov, P.; Hochstrat, A.; Chen, X.; Kleemann, W.; Binek, C. Magnetoelectric Switching of Exchange Bias. *Phys. Rev. Lett.* **2005**, *94* (11), 117203. https://doi.org/10.1103/PhysRevLett.94.117203.

(44) He, X.; Wang, Y.; Wu, N.; Caruso, A. N.; Vescovo, E.; Belashchenko, K. D.; Dowben, P. A.; Binek, C. Robust Isothermal Electric Control of Exchange Bias at Room Temperature. *Nat. Mater.* **2010**, *9* (7), 579–585. https://doi.org/10.1038/nmat2785.

(45) Ashida, T.; Oida, M.; Shimomura, N.; Nozaki, T.; Shibata, T.; Sahashi, M. Observation of Magnetoelectric Effect in $Cr_2O_3$ /Pt/Co Thin Film System. *Appl. Phys. Lett.* **2014**, *104* (15), 152409. https://doi.org/10.1063/1.4871515.

(46) Toyoki, K.; Shiratsuchi, Y.; Nakamura, T.; Mitsumata, C.; Harimoto, S.; Takechi, Y.; Nishimura, T.; Nomura, H.; Nakatani, R. Equilibrium Surface Magnetization of α-$Cr_2O_3$ Studied Through Interfacial Chromium Magnetization in Co/α-$Cr_2O_3$ Layered Structures. *Appl. Phys. Express* **2014**, *7* (11), 114201. https://doi.org/10.7567/APEX.7.114201.

(47) Toyoki, K.; Shiratsuchi, Y.; Kobane, A.; Mitsumata, C.; Kotani, Y.; Nakamura, T.; Nakatani, R. Magnetoelectric Switching of Perpendicular Exchange Bias in Pt/Co/α-$Cr_2O_3$/Pt Stacked Films. *Appl. Phys. Lett.* **2015**, *106* (16). https://doi.org/10.1063/1.4918940.

(48) Shiratsuchi, Y.; Toyoki, K.; Nakatani, R. Magnetoelectric Control of Antiferromagnetic Domain State in $Cr_2O_3$ Thin Film. *J. Phys. Condens. Matter* **2021**, *33* (24), 243001. https://doi.org/10.1088/1361-648X/abf51c.

(49) Bousquet, E.; Lelièvre-Berna, E.; Qureshi, N.; Soh, J.-R.; Spaldin, N. A.; Urru, A.; Verbeek, X. H.; Weber, S. F. On the Sign of the Linear Magnetoelectric Coefficient in $Cr_2O_3$. *J. Phys. Condens. Matter* **2024**, *36* (15), 155701. https://doi.org/10.1088/1361-648X/ad1a59.

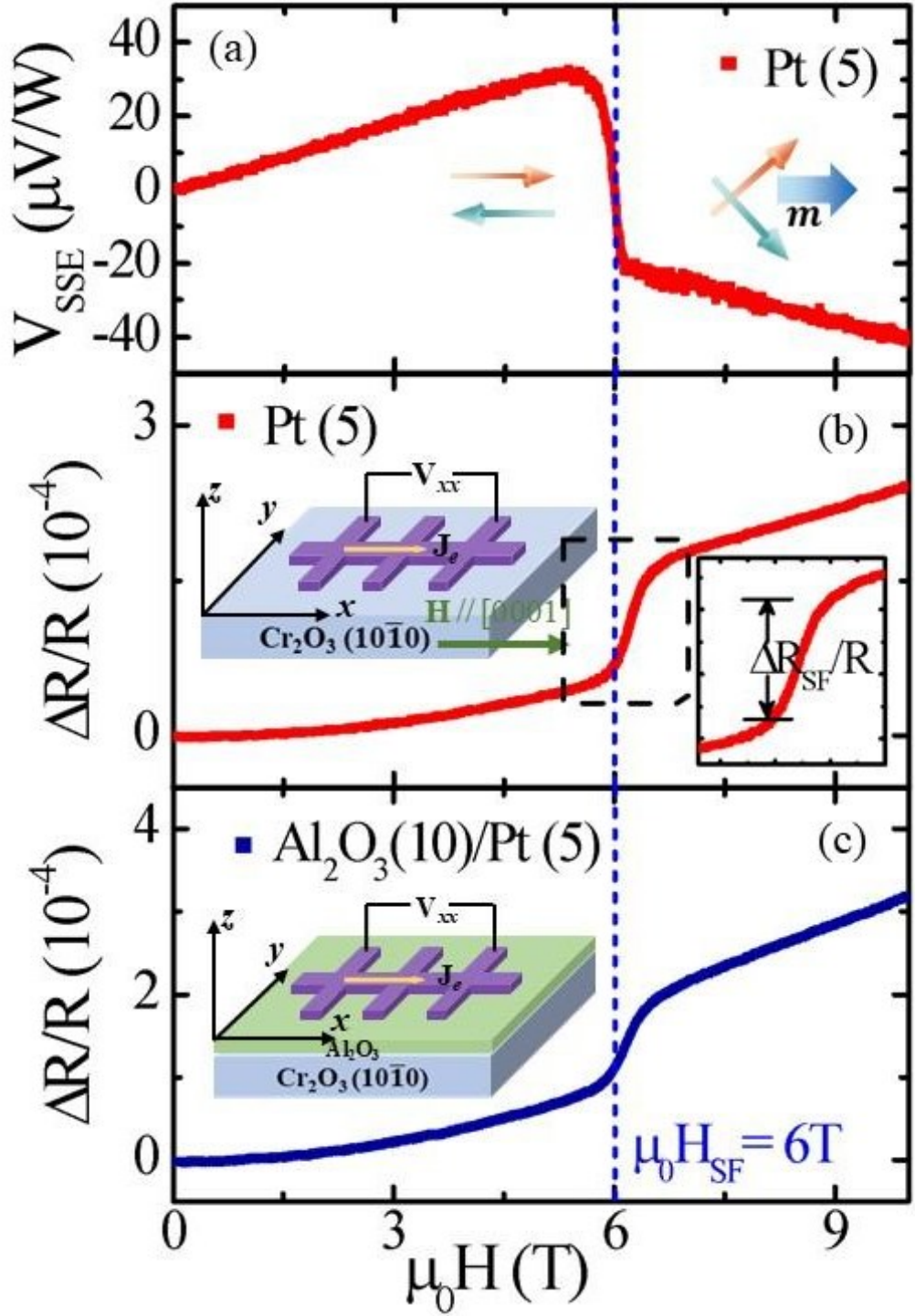

**Figure 1. Spin-flop transition of $Cr_2O_3(10\underline{1}0)$ crystal.** (a) SFT detected by spin Seebeck effect at 50 K. (b)&(c) SFT induced resistivity jump in $Cr_2O_3$/Pt and $Cr_2O_3$/$Al_2O_3$/Pt at 20 K. Insets show the measurement schematics.

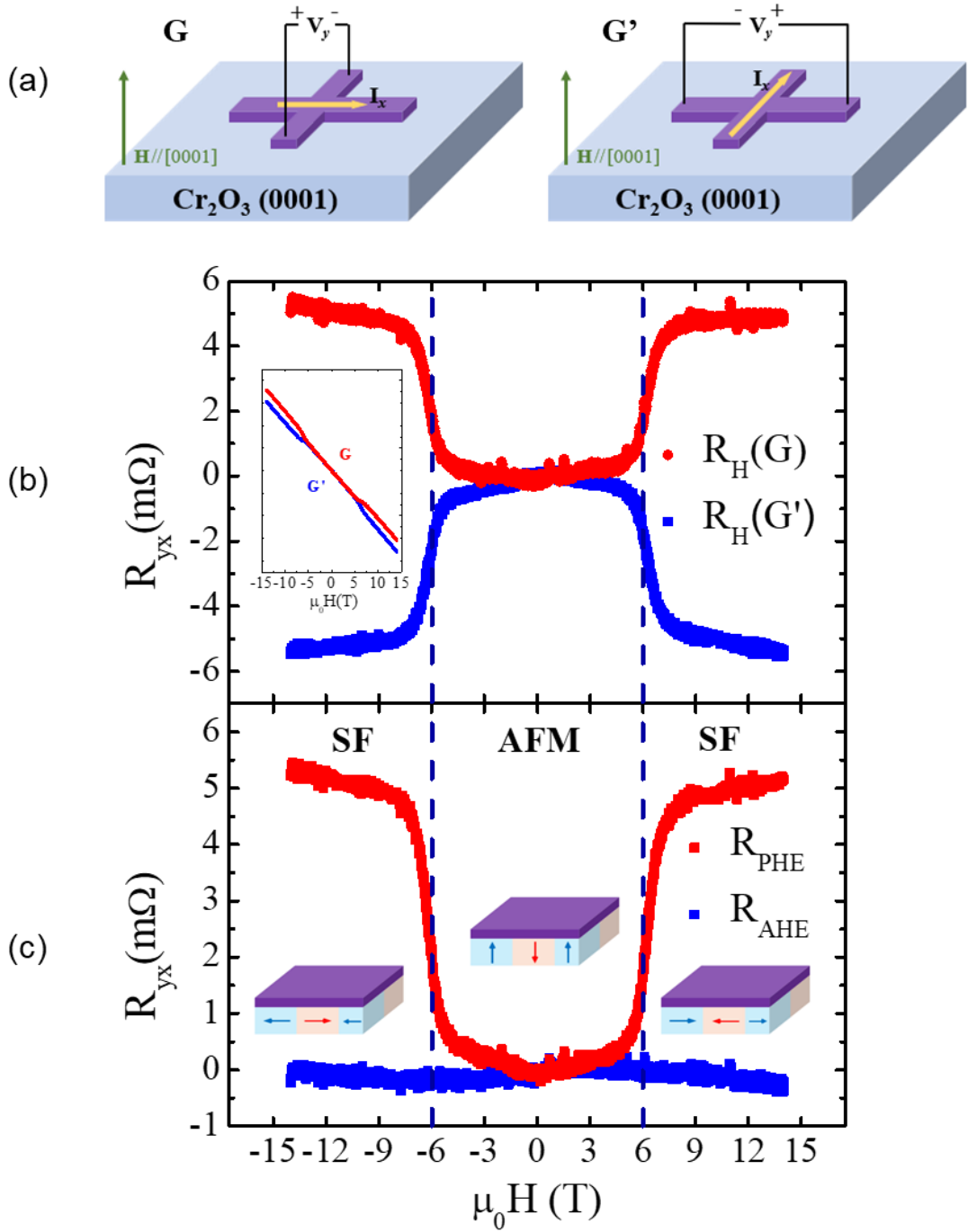

**Figure 2. Hall responses in $Cr_2O_3$(0001)/Pt.** (a) Schematic diagrams of G and G' geometries for Hall measurements. (b) Hall signals $R_{yx}$ measured in G ($R_H$(G)) and G' ($R_H$(G')) at 2 K after the OHE background is removed. The inset shows the raw Hall signals before the linear OHE background removal. (c) $R_{AHE} = (R_H(G)+R_H(G'))/2$ and $R_{PHE} = (R_H(G)-R_H(G'))/2$ are displayed. The cartoons show multi-domain spin configurations in two regimes: "AFM" denoting the state below the AFT, and "SF" denoting the state above.

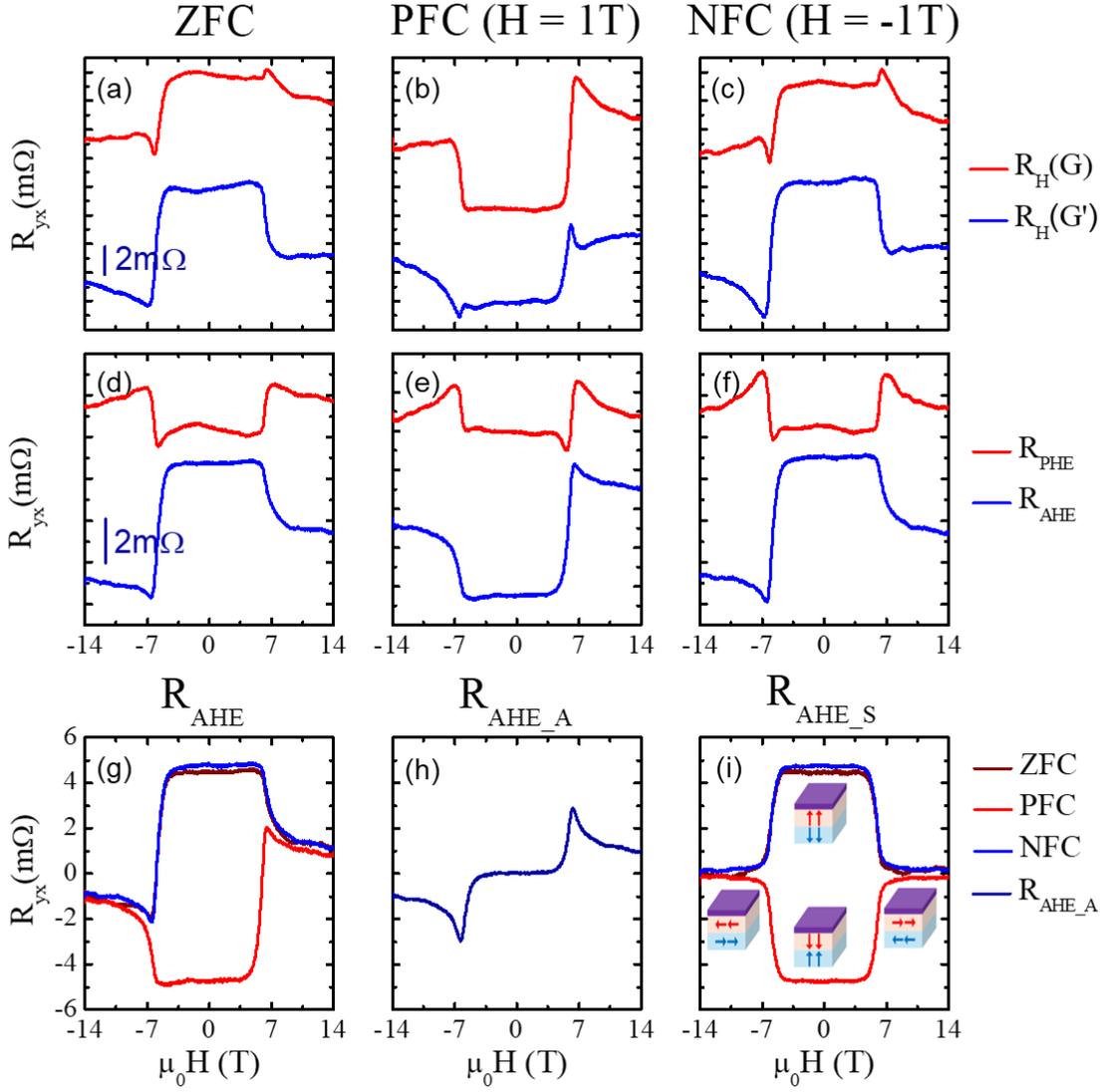

**Figure 3. Hall measurements under different FC conditions.** (a)-(c) $R_H(G)$ & $R_H(G')$, and (d)-(f) $R_{AHE}$ & $R_{PHE}$ for three different FC conditions: ZFC (a, d), PFC (b, e), and NFC (c, f). (g) $R_{AHE}$ for three FC conditions. (h) $R_{AHE\_A}$ derived from averaging $R_{AHE}$(PFC) and $R_{AHE}$(NFC). (i) $R_{AHE\_S}$ for three FC conditions. The cartoons show the corresponding single-domain spin configurations.

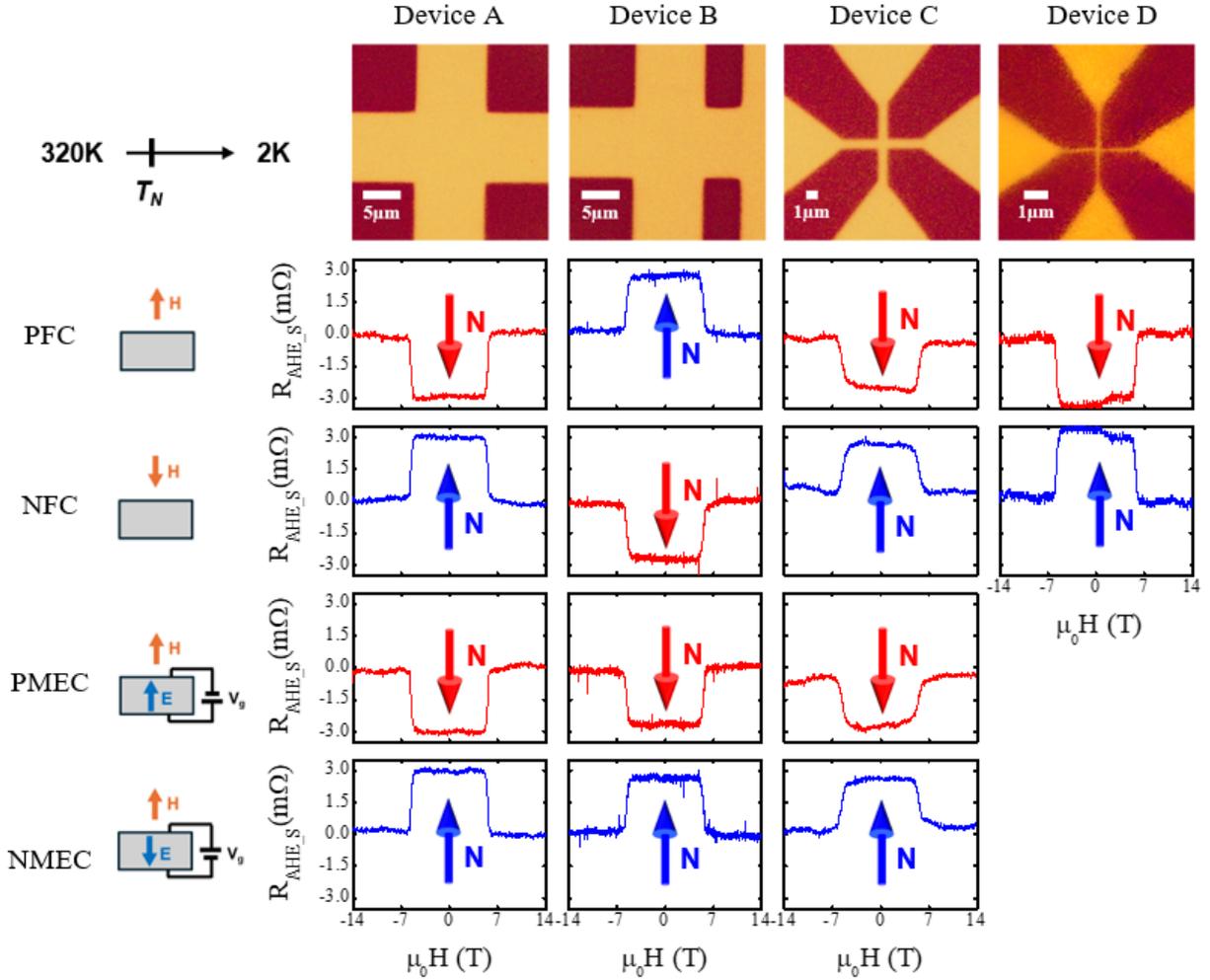

**Figure 4. AHE signals in Pt Hall cross devices prepared with different conditions.** The upper optical micrographs present four Hall cross devices identified as A and B (10 µm), C (1 µm), and D (0.1 µm). $R_{AHE\_S}$ data under PFC and NFC conditions are shown in the first two rows for the four devices. Arrows labeled with "N" represent Néel vector orientations. $R_{AHE\_S}$ data under PMEC and NMEC conditions are shown in the bottom two rows for three tested devices. All Hall cross devices detect single domains. FC is performed from 320 K to 2 K under ±1 T, and MEC is performed under 1T and ±50 V across the 0.5 mm thick $Cr_2O_3$ crystal.

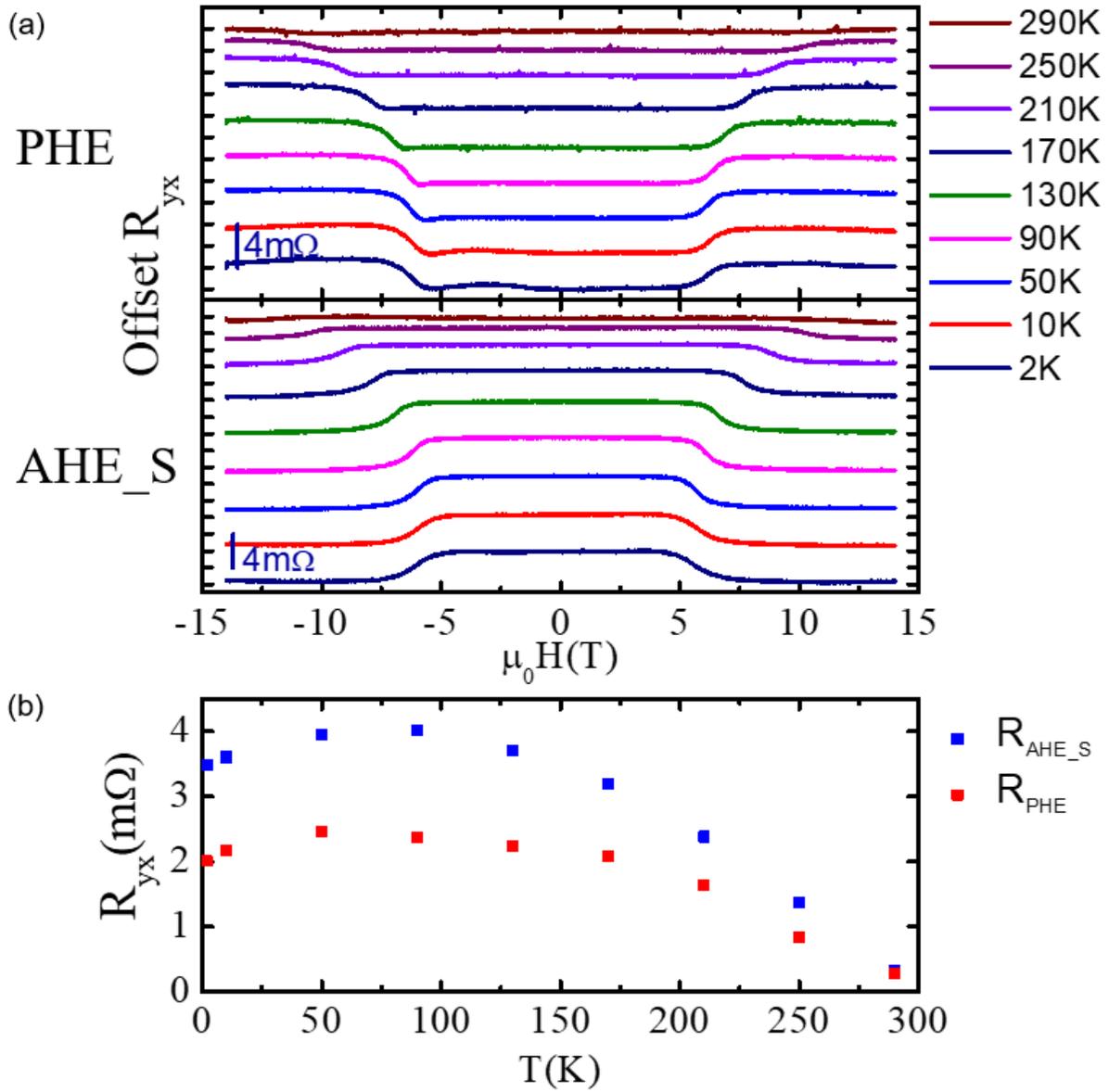

**Figure 5. Temperature Dependence of AHE and PHE signals**. (a) $R_{AHE\_S}$ & $R_{PHE}$ curves at different temperatures. (b) Corresponding $R_{AHE\_S}$ and $R_{PHE}$ magnitudes vs. temperature.